\documentclass[apjl]{emulateapj}
\usepackage{amsmath,amssymb,bm,amstext}
\usepackage{epsfig}
\usepackage{aas_macros}

\begin{document}

\title{Polar alignment of a protoplanetary disk around an eccentric
  binary} \author{Rebecca G. Martin{$^1$} and Stephen H. Lubow{$^2$}}
\affil{ {$^1$}Department of Physics and Astronomy, University of
  Nevada, Las Vegas, Las Vegas, NV 89154, USA\\ {$^2$}Space Telescope
  Science Institute, Baltimore, MD 21218, USA\\ }

\begin{abstract}
We use three-dimensional hydrodynamical simulations to show that an initially  mildly
misaligned circumbinary accretion disk around an eccentric binary can
 evolve to an orientation that is perpendicular to the orbital plane of the binary (polar alignment). 
As the disk evolves to the perpendicular state, it undergoes nodal libration oscillations of the tilt
angle and the longitude of the ascending node. Dissipation within the disk causes the
oscillations to damp.
The process operates
above a critical initial misalignment angle that depends upon the
eccentricity of the binary and the mass of the disk.
For binary eccentricity of 0.5, the process operates typically for disk masses smaller than a few percent of the binary mass and initial tilt angle of more than 40 degrees. This evolution has important implications for planet
formation around eccentric binary star systems.
\end{abstract}


\section{Introduction}

Most stars form in binary or multiple systems
\citep{Ghez1993,Duchene2013} and circumstellar and circumbinary disks
likely form in these systems \citep[e.g.][]{Dutrey1994}.  Stars form in
molecular clouds which are turbulent
\citep[e.g.][]{McKee2007}. This environment may lead to chaotic accretion during
the star formation process \citep{Bate2003}. The result is that
circumbinary disks may form misaligned to the orbital plane of the
binary \citep{Monin2007, Bateetal2010}. Planet formation likely occurs
in these disks and and so understanding the evolution of disks in
these systems is vital for explaining exoplanet properties.

Misalignments have been observed at different evolutionary stages of
binary star systems. The pre--main sequence binary KH 15D has a
circumbinary disk that is inclined and precessing with respected to
the binary orbital plane
\citep[e.g.][]{Winn2004,Chiang2004,Capelo2012}. The circumbinary disk
around the binary protostar IRS~43 has a misalignment of at least
$60^\circ$ \citep{Brinch2016}.  Main--sequence stars in binaries with
separations greater than about $40\,\rm AU$ have spin rotation axes that
are misaligned to their binary orbital rotation axes \citep{Hale1994}. Misaligned
disks around each component of binaries are common
\citep[e.g.][]{Stapelfeldt1998,Jensen2014,Williams2014}. Binary 99 Herculis
 has orbital eccentricity $0.76$ and a misaligned debris disk. The
most likely inclination for the disk is perpendicular to the binary
orbital plane \citep{Kennedy2012}. Finally, exoplanets are observed
with orbits tilted with respect to the spin axis of the central star
\citep{Albrechtetal2012,Winn2015}.

For a small level of misalignment with respect to the binary orbital plane of a circular orbit binary, nearly circular circumbinary test particle orbits undergo gyroscopic motion about the orbital rotation axis of the binary. In this case, particle orbits exhibit nodal precession that is fully circulating with nearly constant tilt. 
But for larger tilts involving eccentric binaries, the  situation is more complicated.
Test particle orbits around a misaligned eccentric binary sometimes undergo nodal
libration (rather than circulation), together with inclination oscillations \citep{Farago2010,Doolin2011}.  The critical misalignment angle
above which this mechanism operates decreases with the eccentricity of
the binary. For circular orbit binaries, the mechanism does not
operate. For low eccentricity, $e=0.2$, the critical inclination is
about $60^\circ$ while for higher eccentricity, $e=0.5$, the critical
angle is about $40^\circ$.

A slightly misaligned circumbinary disk has been predicted to align with the binary
orbital plane through viscous dissipation
\citep[e.g.][]{Nixonetal2011b,Foucart2013,Foucart2014}.  The dynamics
of misaligned circumbinary disks have previously been studied, but the studies
typically assumed that either the binary orbit is circular
\cite[e.g.][]{Nixon2012,Facchinietal2013,Lodato2013,Foucart2013} or
that the binary orbit is eccentric but coplanar to the disk
\citep[e.g.][]{Artymowicz1994, Dunhill2015,Fleming2016}.  \cite{Aly2015} examined the
misaligned and eccentric case for cold black hole disks. They observed
precession about the eccentricity vector of the binary (rather than the orbital rotation axis of the binary), with a cold disk, and aspect ratio
less than the viscosity parameter, $H/R<\alpha$. The disk tore into disjoint
radial regions leading to violent interactions \citep[see
  also][]{Nixonetal2013}. In this work we focus on the wave--like
regime in which $H/R>\alpha$ relevant to protoplanetary disks.

\begin{figure*} 
\begin{centering} 
\includegraphics[width=8.4cm]{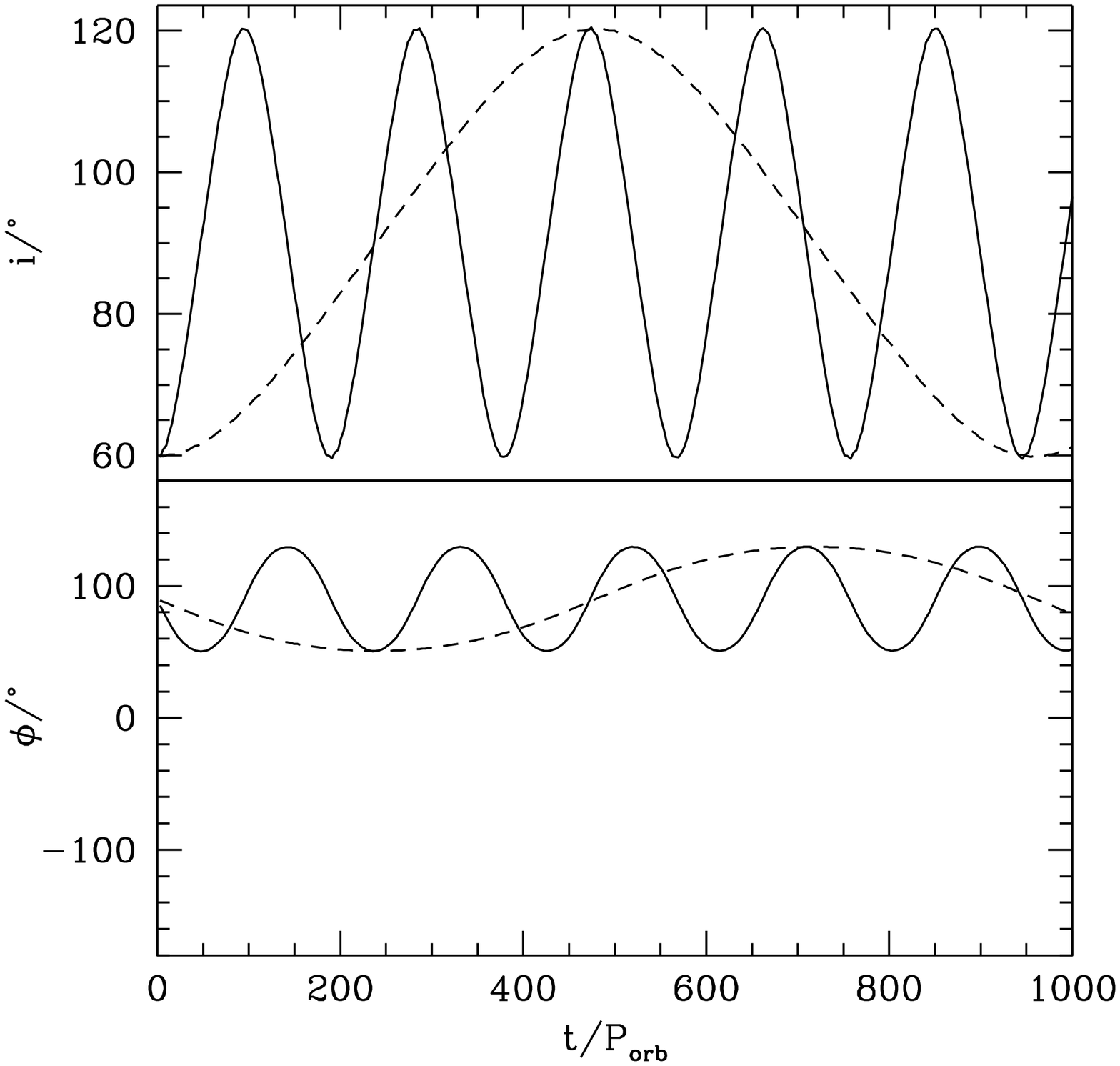}
\includegraphics[width=8.4cm]{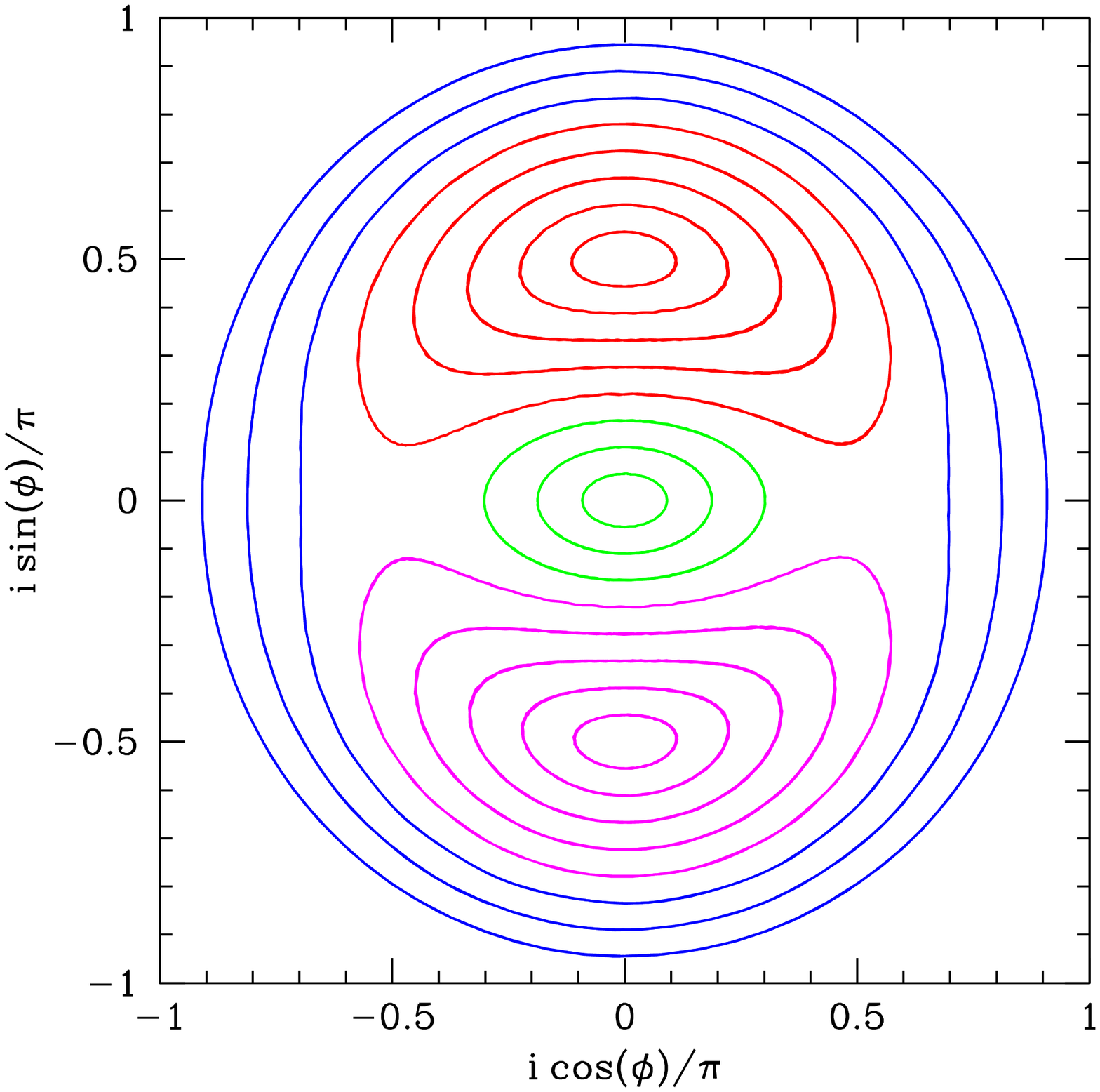}
\caption{Initially circular test particle orbits around an eccentric
  binary with $e=0.5$.  Left: Time evolution of the inclination (upper
  panel) and the longitude of the ascending node (lower panel) for
  orbits initially inclined by $i_0=60^\circ$ to the binary orbit. The
  initial separation is $d=5\,a$ (solid lines) and $d=8\,a$ (dashed
  lines). The initial longitude of the ascending node is
  $\phi_0=90^\circ$.  Right: The $i\cos \phi-i\sin\phi$ plane for
  orbits with varying initial inclination and longitude of the
  ascending node. The green lines show orbits close to prograde with
  $i_0=10^\circ$, $i_0=20^\circ$ and $i_0=30^\circ$ with
  $\phi_0=90^\circ$ in order of increasing size. The blue lines show
  orbits close to retrograde with $i_0=150^\circ$, $i_0=160^\circ$ and
  $i_0=170^\circ$ with $\phi_0=90^\circ$ in order of increasing
  size. The red lines show librating solutions with $i_0=80^\circ$,
  $i_0=70^\circ$, $i_0=60^\circ$, $i_0=50^\circ$ and $i_0=40^\circ$
  with $\phi_0=90^\circ$ in order of increasing size. The magenta
  lines show librating solutions with $i_0=80^\circ$, $i_0=70^\circ$,
  $i_0=60^\circ$, $i_0=50^\circ$ and $i_0=40^\circ$ with
  $\phi_0=-90^\circ$ in order of increasing size. }
\label{particle} 
\end{centering} 
\end{figure*}  

In Section~2 we first reexamine test particle orbits around a
misaligned eccentric binary for parameters relevant to circumbinary
disks around young stars.  In Section~3 we then consider the evolution
of a hydrodynamic protoplanetary disk around a misaligned and
eccentric binary.  In Section~4 we discuss the implications of our
results and we draw our conclusions in Section~5.

\section{Inclined circumbinary test particle obits}

In this Section we consider inclined test particle orbits around an
eccentric binary. The stars have equal mass $M_1=M_2=0.5\,M$, where
$M$ is the total mass of the binary and they orbit with semi--major
axis, $a$.  The eccentricity of the binary is $e=0.5$ and the orbital
period is $P_{\rm orb}=2\pi/\sqrt{G(M_1+M_2)/a^3}$. We work in the frame of
the centre of mass of the binary. In Cartesian coordinates, with the
binary orbit in the $x-y$ plane, the binary begins at periastron
separation on the $x$ axis.

The test particle begins in a circular Keplerian orbit at position
$(0,d,0)$ with velocity $(-\Omega_{\rm p} \cos i_0,0,\Omega_{\rm p}
\sin i_0)$, where $\Omega_{\rm p}=\sqrt{G(M_1+M_2)/d^3}$ is the
Keplerian angular velocity about the centre of mass of the binary and
$i_0$ is the initial particle tilt with respect to the binary orbital plane. The
longitude of the ascending node is measured from the $x$-axis. These
initial conditions correspond to an initial longitude of the ascending
node of $\phi_0=90^\circ$.

The left panel of Fig.~\ref{particle} shows the test particle orbit
evolution for an initial inclination of $i_0=60^\circ$ for two
different initial separations, $d=5\,a$ and $d=8\,a$. The upper panel
shows the inclination of the orbit, $i$, and the lower panel shows the
longitude of the ascending node, $\phi$. The semi-major axis of the
particle remains close to constant over the orbit.  The inclination
and the longitude of the ascending node show synchronous
oscillations. The magnitude of the oscillations does not depend upon
the distance of the particle from the centre of mass of the
binary. However, the timescale for the oscillations increases with
distance.

The right hand panel of Fig.~\ref{particle} shows test particle orbits
in the $i\cos \phi -i \sin \phi$ phase space. The test particles all
begin at a separation of $d=5\,a$, although the separation does not
affect the motion in this phase portrait, only the time taken to make
a complete orbit. Above a certain initial inclination, the particle
orbits undergo libration rather than circulation. The centre of the upper librating region corresponds to $i=90^{\circ}$ and $\phi=90^{\circ}$, while the centre of the lower librating region corresponds to $i=90^{\circ}$ and $\phi=-90^{\circ}$. 
For higher binary eccentricity, the critical inclination tilt angle that separates
the librating from  circulating cases is
smaller \citep[see][for more details]{Doolin2011}. When the third body is massive, the nodal libration region shrinks \citep[see Fig.~5 in][]{Farago2010}.  For a body with a mass of the order of a few percent of the binary mass, the region may be reduced somewhat for the configuration of bodies considered here. In the next Section
we consider the evolution of a misaligned low mass circumbinary disk around an
eccentric binary.

\section{Circumbinary disk Simulations}

\begin{table}
\caption{Parameters of the initial circumbinary disk set up for an
  eccentric, equal mass binary with total mass, $M$, and separation,
  $a$.} \centering
\begin{tabular}{lllll}
\hline
Binary and Disk Parameters & Symbol & Value \\
\hline
\hline
Mass of each binary component &  $M_1/M = M_2/M$ & 0.5 \\
Eccentricity of the binary & $e$ & 0.5 \\
Accretion radius of the masses & $R_{\rm acc}/a$    & 0.25  \\
Initial  disk mass & $M_{\rm di}/M$ & 0.001 \\
Initial disk inner radius & $R_{\rm in}/a$ & 2 \\
Initial disk outer radius & $R_{\rm out}/a$ & 5 \\
Disk viscosity parameter & $\alpha$ & $0.01$ \\
Disk aspect ratio & $H/R (R=R_{\rm in})$ & 0.1 \\
   & $H/R (R=R_{\rm out})$ & 0.08 \\
Initial disk inclination & $i$ & $60^\circ$ \\ 
\hline
\end{tabular}
\label{tab}
\end{table}

\begin{figure*} 
\begin{centering} 
\includegraphics[width=16cm]{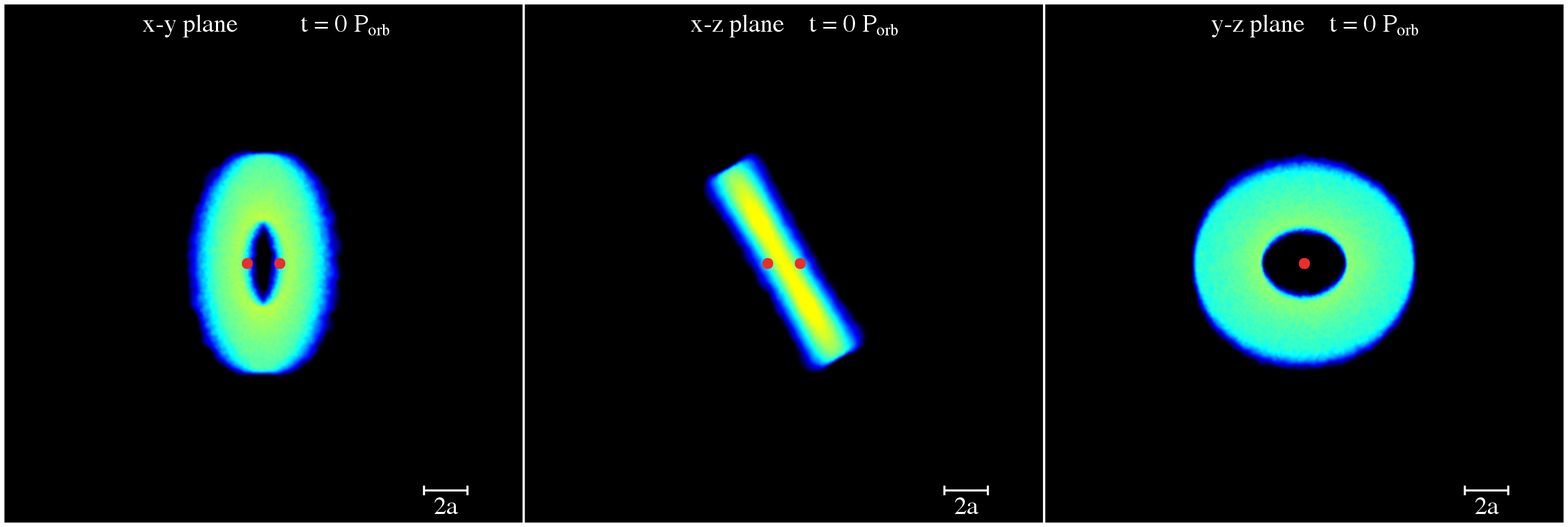}
\includegraphics[width=16cm]{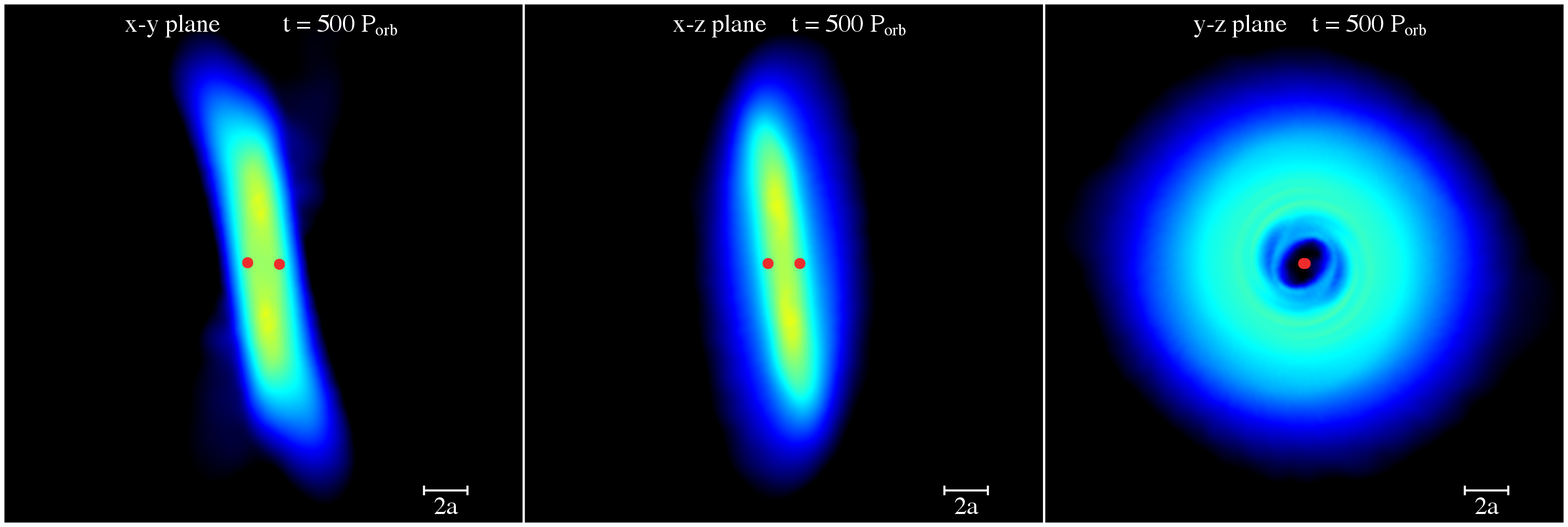}
\caption{Upper panels: Initial disk set up for the SPH simulation of a
  binary (shown by the red circles) with an inclined circumbinary
  disk. Lower panels: The disk at a time of $t=500\, P_{\rm orb}$. The
  color denotes the gas density with yellow regions being about two
  orders of magnitude larger than the blue. The left panels show the
  view looking down on to the binary orbital plane, the $x-y$
  plane. The middle panels show the $x-z$ plane and the right panels
  show the $y-z$ plane. In the right hand panels the binary components
  lie in front of each other and so only one red point is seen. }
\label{discpics} 
\end{centering} 
\end{figure*} 

\begin{figure*} 
\begin{centering} 
\includegraphics[width=8.4cm]{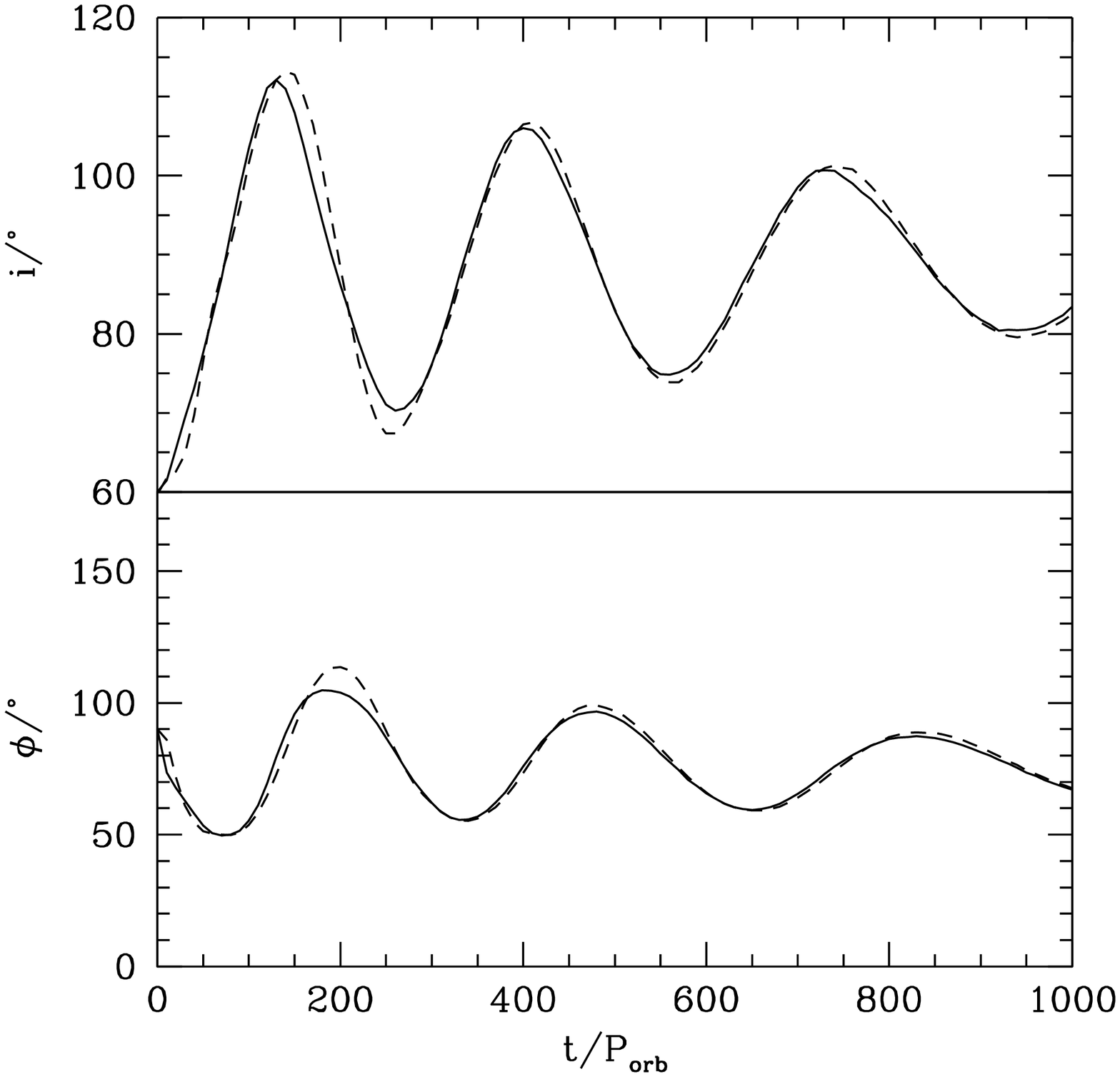}
\includegraphics[width=8.4cm]{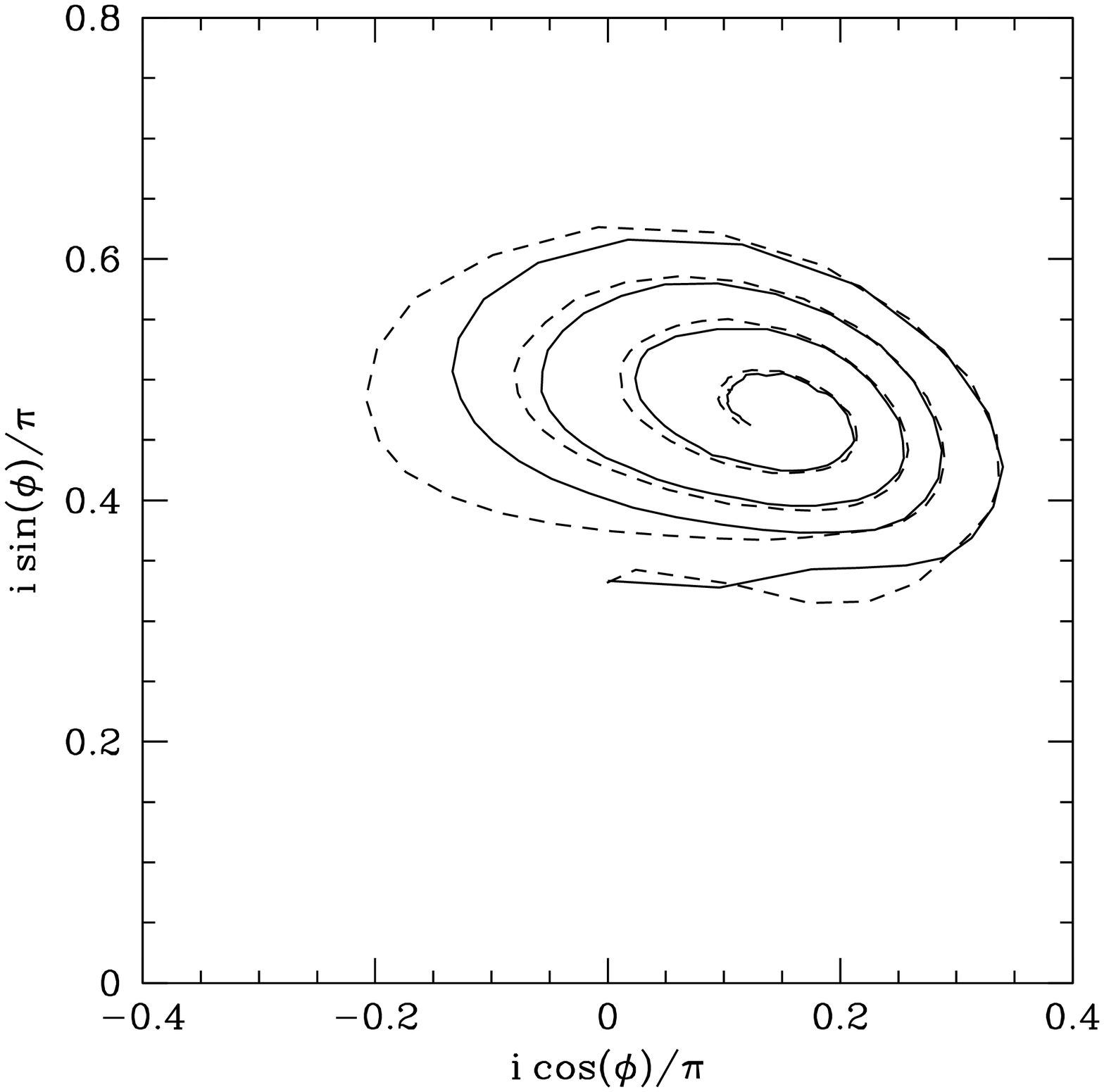}
\caption{Left panel: Inclination (upper panel) and longitude of the
  ascending node (lower panel) for a circumbinary disk that is
  initially misaligned by $i_0=60^\circ$.  The solid lines show the
  disk evolution at a radius of $3\,a$ and the dashed lines show a
  radius of $5\,a$. Right panel: The same simulation in the $i\cos
  \phi-i\sin\phi$ plane at a radius of $3\, a$ (solid line) and $5\,a$
  (dashed line).}
\label{disc} 
\end{centering} 
\end{figure*}

With hydrodynamic disk simulations we now analyze the evolution of a
misaligned circumbinary disk around an eccentric equal mass binary.
We use the smoothed particle hydrodynamics (SPH;
e.g. \citealt{Price2012a}) code {\sc phantom} \citep{PF2010,LP2010}.
{\sc phantom} has been used to model misaligned accretion disks in
binary systems previously
\citep[e.g.][]{Nixon2012,Nixonetal2013,Martinetal2014b,Fu2015}
Table~\ref{tab} contains a summary of the binary and disk parameters.  The
binary has equal mass components with total mass $M=M_1+M_2$, and an eccentric orbit in the $x$-$y$ plane with semi--major axis, $a$. The
accretion radius for particle removal from the simulation about each
star is $0.25\,a$.

The upper panels in Fig.~\ref{discpics} show the initially flat and
circular disk that is tilted to the binary orbital plane by
$60^\circ$. The disk has initial mass $10^{-3}\,M$ distributed in $300,000$ equal mass
particles.  The small disk mass has a minor dynamical significance on the
orbit of the binary. Disk self-gravity is ignored. The initial
surface density profile has a power law distribution $\Sigma \propto
R^{-3/2}$ between $R_{\rm in}=2\, a$ and $R_{\rm out}=5\,a$. The inner
radius of the disk is chosen to be close to the radius where a disk is
tidally truncated \citep{Artymowicz1994}. However, misaligned disks
feel a weaker binary torque
\citep[e.g.][]{Lubowetal2015,Nixon2015,Miranda2015}.  The disk is
locally isothermal with sound speed $c_{\rm s} \propto R^{-3/4}$ and
$H/R=0.1$ at $R=R_{\rm in}$. This choice allows both $\alpha$ and
$\left<h \right>/H$ to be constant over the radial extent of the disk
\citep{LP2007}. The \cite{SS1973} $\alpha$ parameter is taken to be
0.01 (the disk viscosity is implemented in the usual manner by
adapting the SPH artificial viscosity according to \cite{LP2010} with
$\alpha_{\rm AV} = 0.4$ and $\beta_{\rm AV} = 2.0$). The disk is
resolved with shell-averaged smoothing length per scale height
$\left<h\right> /H \approx 0.25$.

In the left hand panel of Fig.~\ref{disc} we show the time evolution
of the inclination and the longitude of the ascending node of the disk
at two orbital radii, $d=3\,a$ and $d=5\,a$. We clearly see damped
nodal libration of the disk.  As the inclination increases, the
longitude of the ascending node decreases and vice versa. The
evolution is very similar for different radii in the disk since the
disk radial communication timescale is short enough for the nodal libration
to occur globally. Dissipation  causes the disk to move towards an
inclination perpendicular to the binary orbit. The right hand panel
shows the spiral in the $i\cos \phi-i\sin \phi$ plane as the disk
tilt evolves towards being perpendicular to the binary orbital plane at the center
of the librating region. We note that the spiral is slightly offset to the right of the diagram compared with the test particle orbits. This offset is due to the apsidal precession of the binary during the simulation due to the nonzero disk mass. This does not occur in the test particle case because the binary is not affected by the particle. The disk moves towards polar  alignment perpendicular to the eccentricity vector of the binary. In the lower panels of Fig.~\ref{discpics}, we show the disk at a time of $t=500\,P_{\rm b}$
when the disk is almost perpendicular to the binary orbital plane.

\section{Discussion}

Although we do not present the results here, we have also examined some
different parameters for simulations. First, we have considered
disks with a larger radial extent. We find for a disk initially outer
truncated at larger radius that the evolution is at least initially
qualitatively the same. The disk displays oscillations and moves
towards a perpendicular orientation. The oscillations are more strongly damped for a disk
with a larger radial extent. However, for disks truncated at radius
$\gtrsim 10\,a$, there is some damping of the binary eccentricity. Lower binary eccentricity reduces  the tendency for polar alignment.
There is thus a competition between the timescales for the binary eccentricity
damping and the polar alignment. Furthermore, if the alignment timescale
becomes shorter than the sound crossing timescale, then  disk warping will occur.

We have run the same simulation that is presented here,
but with a circular binary and find that the alignment proceeds
towards the binary orbital plane. We have also explored the
evolution of a disk that begins close to counter alignment and find
that the disk moves closer to counter--alignment. We have also
considered the effect of a larger disk mass. We find that the
accretion of material from a disk of mass $0.05\,M$ can circularise
the binary. Furthermore, for large disk masses, the apsidal precession timescale of the binary may become shorter than the libration timescale of the disk, in which case the disk more closely follows a circulating solution. Both of these effects, accretion and precession, can result in disk--binary planar alignment, rather than polar alignment.  If we observe a disk, or a planet, to be in a polar orbit,
the eccentricity of the binary  places constraints on 
the mass of the circumbinary disk.  These effects will all be
investigated in future work.

The circumbinary disk around the binary in KH 15D may be subject to the nodal and tilt oscillations described here. The binary
eccentricity is high, $0.68<e<0.8$ \citep{Johnson2004}. For an
eccentricity of 0.8, the mechanism described in this work would polar
align the disk for a modest initial inclination of $20^\circ$ and a low disk mass
\citep[see][]{Doolin2011}. A narrow ring disk has been invoked to
explain the peculiar light curve. Thus the disk may be
precessing about the eccentricity vector of the binary rather than the binary orbital
axis. Over time, the disk will align with the polar axis rather than
the binary orbital axis.

\cite{Kennedy2012} pointed out that the debris disk observed in 99
Herculis
 could be polar due to the stability of perpendicular particle orbits in this highly eccentric binary.
In this model, the particles represent the solid debris.
The results of this work suggest that the debris disk observed in 99
Herculis most likely arose from the evolution of a misaligned protostellar disk
surrounding an eccentric binary.  
 Because the eccentricity of the binary is
very high at 0.76, the initial misalignment would not have to be very
large.  Material misaligned by only $20^\circ$ to the binary orbital
plane would evolve to become perpendicular. Thus, in binaries with
large eccentricities, perpendicular disks may be more likely than
coplanar disks. Debris disks and any planets that form in the disk may
be more likely to be polar with orbital axes parallel to the binary eccentricity vector rather than aligned with the binary.

\section{Conclusions}

We have found that a polar alignment mechanism can operate for
inclined disks around an eccentric binary star system. The mechanism operates best for higher binary eccentricity, larger initial disk misalignment angle,  and lower disk mass.
The inclination
of the disk is exchanged with the longitude of the ascending
node. Dissipation within the disk aligns the disk to be perpendicular to
the binary orbital plane with disk rotation axis parallel to the binary eccentricity vector. The results have many implications for circumbinary gas
disks, circumbinary planets, and circumbinary debris disks.

\acknowledgements

S.H.L. acknowledges support from NASA grant NNX11AK61G.  Computing
resources supporting this work were provided by the UNLV National
Supercomputing Institute. We thank Daniel Price for providing the {\sc
  phantom} code for SPH simulations and the {\sc splash} code
\citep{Price2007} for data analysis and rendering of figures.

\bibliographystyle{apj}

\end{document}